\documentclass[sigconf,authorversion,nonacm]{acmart}

\AtBeginDocument{%
  }

\setcopyright{acmlicensed}
\copyrightyear{2018}
\acmYear{2018}
\acmDOI{XXXXXXX.XXXXXXX}

\acmConference[Conference acronym 'XX]{Make sure to enter the correct
  conference title from your rights confirmation emai}{June 03--05,
  2018}{Woodstock, NY}
\acmISBN{978-1-4503-XXXX-X/18/06}

\begin{document}
\title{StreamLink: Large-Language-Model Driven Distributed Data Engineering System}

\author{Dawei Feng}
\affiliation{%
  \institution{Tsinghua University}
  \city{Beijing}
  \country{China}
}
\email{fdw22@mails.tsinghua.edu.cn}

\author{Di Mei}
\affiliation{%
  \institution{Tsinghua University}
  \city{Beijing}
  \country{China}
}
\email{di.mei@rioslab.org}

\author{Huiri Tan}
\affiliation{%
  \institution{Tsinghua University}
  \city{Beijing}
  \country{China}
}
\email{huiri.tan@rioslab.org}

\author{Lei Ren}
\affiliation{%
  \institution{Tsinghua University}
  \city{Beijing}
  \country{China}
}

\author{Xianying Lou}
\affiliation{%
  \institution{King \& Wood Mallesons}
  \city{Shanghai}
  \country{China}
}

\author{Zhangxi Tan}
\affiliation{%
  \institution{Tsinghua University}
  \city{Beijing}
  \country{China}
}

\begin{abstract}
Large Language Models (LLMs) have shown remarkable proficiency in natural language understanding (NLU)\cite{brown2020language}, opening doors for innovative applications. We introduce StreamLink - an LLM-driven distributed data system designed to improve the efficiency and accessibility of data engineering tasks. We build StreamLink on top of distributed frameworks such as Apache Spark\cite{zaharia2010spark} and Hadoop to handle large data at scale. One of the important design philosophies of StreamLink is to respect user data privacy by utilizing local fine-tuned LLMs instead of a public AI service like ChatGPT. With help from domain-adapted LLMs, we can improve our system's understanding of natural language queries from users in various scenarios and simplify the procedure of generating database queries like the Structured Query Language (SQL) for information processing. We also incorporate LLM-based syntax and security checkers to guarantee the reliability and safety of each generated query. StreamLink illustrates the potential of merging generative LLMs with distributed data processing for comprehensive and user-centric data engineering. With this architecture, we allow users to interact with complex database systems at different scales in a user-friendly and security-ensured manner, where the SQL generation reaches over 10\% of execution accuracy compared to baseline methods, and allow users to find the most concerned item from hundreds of millions of items within a few seconds using natural language.
\end{abstract}

  
\maketitle

\section{Introduction}

Big data is now a key focus for both government and business leaders.\cite{chen2014data}. However, buried within this immense data deluge lies an abundance of untapped potential and valuable insights, which has given rise to an innovative scientific paradigm known as data-intensive scientific discovery\cite{kelling2009data}. Researchers actively seek ways to leverage available data to gain valuable insights and inform decision-making. On the one hand, big data offers substantial value, fostering business productivity and catalyzing revolutionary breakthroughs in scientific disciplines. On the other hand, the utilization of big data is accompanied by challenges, ranging from the complexities of data capture\cite{stone2007science}, storage\cite{siddiqa2017big}, and analysis to the intricacies of data visualization\cite{ali2016big}.



The prerequisite for realizing big data applications is a robust data system managing external queries as well as information retrieval. In order to efficiently and securely handle a large volume of data, we introduce StreamLink, an AI-powered distributed data system with enhanced ability to process billions of data records while reducing user operational costs. In addition to the support from a scalable and dependable distributed database, one exceptional feature of this system is its highly accessible and security-oriented interaction with users, which is contributed by the use of the latest Large Language Models (LLMs) with the outstanding capability of language generation and domain adaptation. To illustrate its performance, we deploy the proposed system to store global patent data, where most of them come from the United States Patent and Trademark Office (USPTO)\footnote{https://www.uspto.gov} and Google Patents\footnote{https://patents.google.com}. There are approximately 180 million patents in this system and patents are growing rapidly, with the USPTO Patent Assignment Dataset\cite{marco2015uspto} containing around 6 million assignments and transactions recorded between 1970 and 2014, affecting about 10 million patents or patent applications. We have also validated Streamlink's robustness through the collaboration with the patent and intellectual property (IP) team at King \& Wood Mallesons, who is invited to experience our system and provide any usage feedback.


In this paper, we make the following contributions:

\begin{figure*}[tp]
  \centering
  \includegraphics[width=\textwidth]{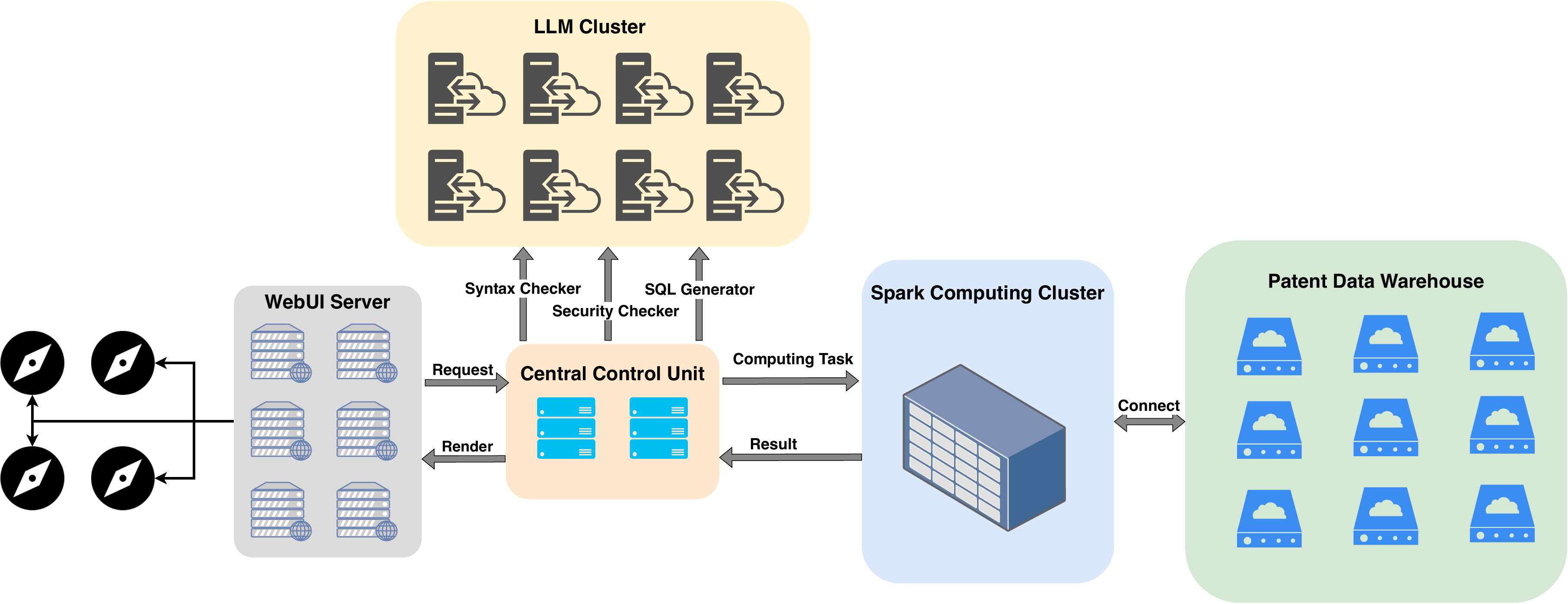}
  \caption{Architecture of Project StreamLink}
  \label{fig:arch}
\end{figure*}

\begin{itemize}
    \item \textbf{LLM-based NL to SQL\cite{jamison2003structured} Generator}: While numerous studies\cite{li2023graphix}\cite{zeng2023n}\cite{zhao2022importance} have explored Natural Language to Structured Query Language (NL-to-SQL) techniques, we integrate the latest advancements in LLM into our distributed patent database system. Our approach leverages LLMs' in-context learning capabilities and domain-specific adaptability to understand and translate natural language instructions into SQL commands that can precisely operate over 180 million patents in our database.

    \item \textbf{Optimized Distributed System Architecture for AI+ Platforms}: Our research focuses on creating scalable AI+ platforms using optimized distributed system architecture. In the traditional Apache distributed architecture, we have added distributed LLM clusters and UI clusters, and further designed a Central Control Unit (CCU) to schedule tasks. We utilize three distributed storage nodes, two LLM nodes, and two UI nodes to conduct tests with 180 million patents. The storage consumption is 15.3TB (5.1TB x 3, with dual redundancy for reliability), and the system is accelerated using 280 cores (560 threads) and 2.6TB of memory. During testing, we confirmed that the average time from user input in natural language to obtaining the desired patent from a database containing over 180 million patents is within 6 seconds.


    \item \textbf{Data Privacy Protecting}: We have confirmed that building a localized LLM-based assistant can significantly enhance productivity while providing a higher level of privacy to users. Through using a locally deployed model for our LLM-based assistant, we can effectively eliminate the risks of data breaches and information leakage which could occur while using cloud-based AI assistants. This approach maintains the confidentiality and integrity of user data and underscores our commitment to prioritizing privacy in the development of advanced technological solutions. We have also developed mechanisms for SQL legality checking using Llama\cite{touvron2023llama} based tools, protecting the system from accidental deletion or injection attacks.
\end{itemize}

We organized this paper as follows. We will discuss the necessity of this work and its application scenarios in Section 2, then introduce the architecture of StreamLink in Section 3, the methodologies we used, and the reason we chose these technologies. We present some experiments in Section 4, including comparisons between our method and traditional strategies with statistical metrics, and conclude the paper with a short discussion in Section 5.

\section{Task Description}
In this section, we will discuss the reason we created the StreamLink project (as shown in Figure \ref{fig:arch}) and provide some cases where it has been used successfully.

Our data system is primarily used for handling large-scale data storage and retrieval. A typical scenario involves retrieving several patents from a database such as Google Patents that meet specific criteria (e.g. date range, pattern, keywords, etc) and analyzing the potential for IP infringement. Traditionally, IP researchers and lawyers might need to read through extensive documentation and perform dozens to hundreds of repetitive searches on Google Patents, or write complex SQL-like statements on data engineering platforms like BigQuery to retrieve patents. The former requires significant manpower and time, often necessitating several lawyers to collaborate over several days to filter the data, while the latter requires extensive technical expertise to write complex SQL commands or other data manipulation languages, and familiarity with the intricacies of data storage and computation frameworks. In addition, SQL commands could have bugs and often require considerable time to be adjusted and modified.

With the StreamLink platform, users can complete all the above tasks in a more efficient and accessible fashion. Without the need to design a SQL command, users like IP researchers and lawyers can directly query the patent database via a natural language request. With the LLM-based interface, our data system converts this natural language query into a SQL command with a security check, and then the distributed database finishes executing this SQL command in a few seconds. Retrieved patents are expected to meet all filter conditions in the natural language query. Furthermore, there is great flexibility for creating different AI interfaces upon StreamLink's distributed database. In this case, we have implemented a BERT-based\cite{reimers2019sentence} semantic filter upon theses retrieved patents to further extract the patents with the potential for IP infringement. 


Another challenge is the scalability of large-scale database. Traditional data warehouses are struggling to handle the exponential growth of data volumes, which can lead to capacity issues\cite{huertas2007layout}, and they can also fail to seamlessly scale in response to fluctuating data processing demands\cite{silva2016sql}. To handle the issue of patent storage capacity, we employed distributed data warehouses\cite{hadoop}, designed to efficiently store and manage a vast amount of information across multiple servers. This ensures high fault tolerance of databases as well as facilitates elastic scaling of storage resources to meet growing patent data demands. Currently, we use three of the 5.1TB nodes to store 180 million entries from the USPTO and Google Patents.

\section{Methodology}

In this section, we will present the components in StreamLink. Section 3.1 will introduce the LLM-driven SQL Generator, an innovative tool capable of understanding natural language instructions and translating them into SQL commands based on the database schema. Section 3.2 will showcase our distributed framework based on Apache Spark and Apache Hadoop, which offer robust support for processing large-scale datasets, ensuring high scalability and processing capacity. Moreover, we will discuss our distributed WebUI clusters and load balancing in this section. We will also talk about our brand new Llama-based SQL syntax and security checker built upon StreamLink to reduce the risks associated with grammatical errors or malicious SQL injections in Section 3.3.

\subsection{LLM-based SQL Generator}

Our IP lawyer collaborators work with various patents from the globe every day. They may want to execute a command similar to \texttt{SELECT cpc, COUNT(*) AS count FROM google\_full WHERE assignee LIKE "\%Intel\%" AND grant\_date >= "2009" GROUP BY cpc ORDER BY count DESC LIMIT 10} to conduct an analysis on the most popular CPC numbers of patents from Intel, but writing such a SQL command is too difficult for them without professional programming training.

To solve this problem, our LLM-driven SQL Generator is an innovative tool that makes data engineering more accessible to a wider audience. It has the ability to comprehend natural language instructions and convert them into SQL commands, thereby reducing the learning curve for users. Even those who lack specialized programming training can effortlessly carry out complex data engineering tasks.


While traditional natural language to SQL generators are based on Encoder and Decoder structures\cite{brunner2021valuenet}, requiring extensive data training to obtain the ability to generate SQL commands before specializing in a specific database, we utilize an LLM-based SQL generator and propose two methods for SQL generation. One method involves quickly generating specialized SQL commands for corresponding databases based on specific rules, followed by fine-tuning. The other method involves parsing database structures to quickly generate prompt templates, aiding LLM in migrating to new databases. Both methods are faster and more scalable than traditional approaches, making them become more appropriate data engineering assistants.

We use LoRA\cite{hu2021lora} as an improved fine-tuning method where instead of fine-tuning all the weights that constitute the weight matrix of the pre-trained LLM, two smaller matrices that approximate this larger matrix's weight update are fine-tuned. These matrices constitute the LoRA adapter. This fine-tuned adapter is then loaded to the pre-trained model and used for inference.

For the NL-to-SQL conversion, we construct context-target pairs: $Z = \{(x_i, y_i)\}_{i=1}^{N}$, where $x_i$ is a natural language query and $y_i$ its correspoding SQL command. During fine-tuning, the model is initialized to pre-trained weights $\Phi_0$, and the task-specific parameter increment $\Delta\Phi = \Delta\Phi(\Theta)$ is further encoded by a much smaller-sized set of parameters $\Theta$ with $|\Theta| \ll |\Phi_0|$. To optimize the SQL generation quality is to minimize the cross-entropy loss at the decoding stage. The task of finding $\Delta\Phi$ thus becomes optimizing over $\Theta$:

\begin{align}
    \max_{\Theta} \sum_{(x,y) \in Z} \sum_{t=1}^{|y|} \log (p_{\Phi_0+\Delta\Phi(\Theta)}(y_t | x, y_{<t}))
\end{align}

Instead of full fine-tuning:

\begin{align}
    \max_{\Phi} \sum_{(x,y) \in Z} \sum_{t=1}^{|y|} \log (P_{\Phi}(y_t|x, y_{<t}))
\end{align}

\begin{figure}[htbp]
  \centering
  \includegraphics[width=\linewidth]{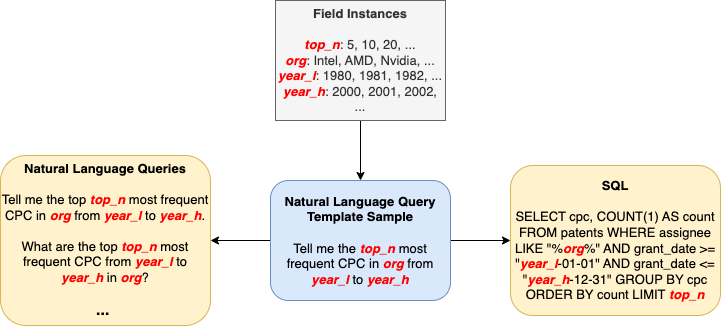}
  \caption{Data augmentation via bi-directional propagation}
  \label{fig:nl2sql_aug}
\end{figure}

Another critical challenge for fine-tuning is to adapt an LLM to NL-2-SQL tasks within a domain-specific schema. Different domains have different rules of defining schemas in their data storage, and thus we proposed a mechanism to augment the domain-specific NL-2-SQL training set given a small set of query templates. This mechanism augments the training set by simultaneously propagating SQL commands and their corresponding natural language queries (see Figure \ref{fig:nl2sql_aug}). Every SQL template query can be turned into a set of SQL commands by inserting different field instances into it; and for each SQL template query, we designed natural language queries in different written expressions. Each SQL template is propagated in two directions (natural language queries and SQL commands with various field instances) and then natural language queries are matched with their corresponding SQL commands to form the augmented training set. To prevent the LLM from suffering from catastrophic forgetting and over-fitting, we combined the domain-specific dataset with publicly available NL-2-SQL datasets like WikiSQL\cite{zhongSeq2SQL2017} and Spider\cite{yu2018spider}. Through extensive experiments, 1:1 is found to be the optimal hybrid ratio of domain-specific training set to the open domains.

\subsection{Distributed Computing, Storage and WebUI Cluster}
In this section, we will discuss our distributed framework. This framework is the foundation of our data engineering system, and it is designed to manage and process large-scale datasets efficiently, making our system scalable and robust. Using these distributed computing paradigms, we can distribute data processing tasks among multiple nodes, reducing the time required for data processing and analysis.

\begin{figure}[htbp]
  \centering
  \includegraphics[width=\linewidth]{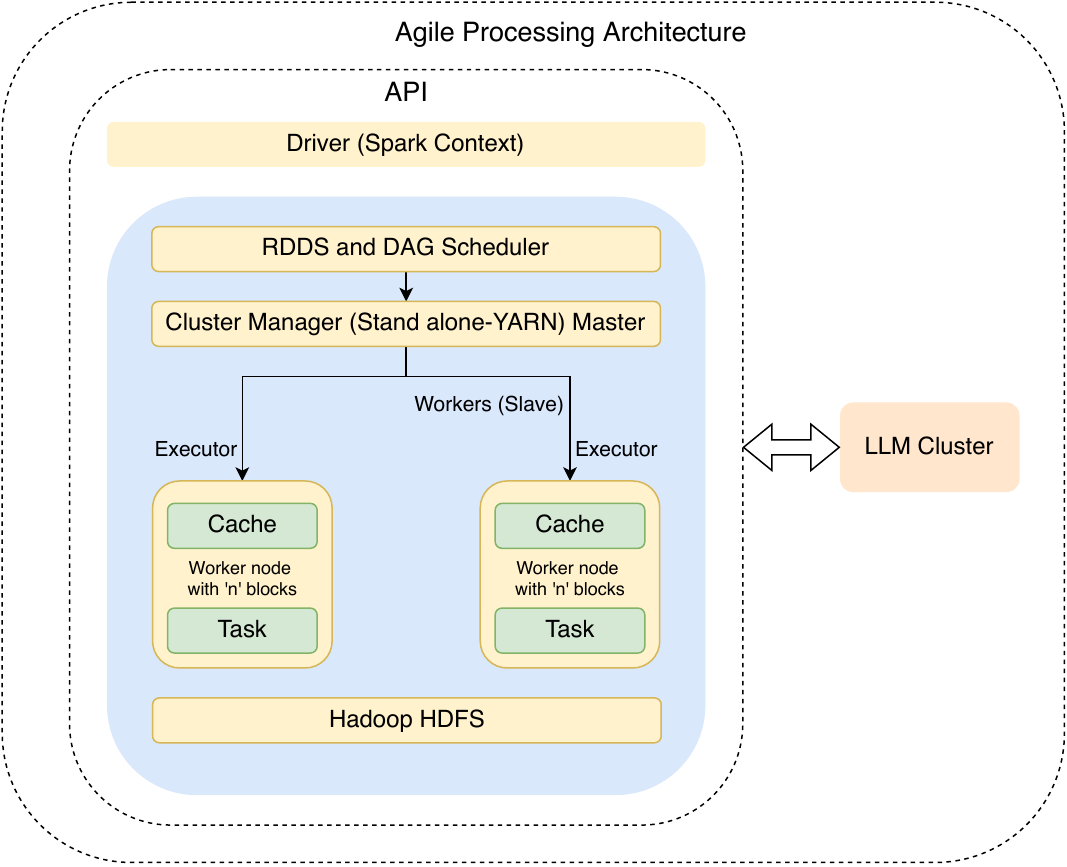}
  \caption{Distrubuted system architecture with LLM to improve agility}
  \label{fig:spark-hadoop}
\end{figure}

As shown in Figure \ref{fig:spark-hadoop}, by adopting this approach, we can efficiently, reliably, and scalably handle large-scale datasets. Not only does this method overcome the limitations of traditional data processing methods, but it also unlocks new possibilities for advanced data analytics and engineering tasks. Therefore, it is an essential component of our data engineering ecosystem.



For user experience, we have developed a distributed Web User Interface (WebUI) cluster and implemented a load balancing mechanism that makes sure high availability and responsiveness of the user interface. To guarantee the effectiveness of our WebUI cluster, we have implemented a robust load balancing mechanism using Nginx\cite{reese2008nginx}, a high-performance HTTP server and reverse proxy. Nginx acts as an intermediary between the client and the WebUI instances, intelligently distributing incoming requests across the available nodes based on predefined algorithms. This evenly distributes incoming traffic across the WebUI instances, preventing any single node from becoming overwhelmed with requests, thus avoiding performance degradation and downtime. Additionally, in case of node failure or maintenance, Nginx dynamically reroutes the requests to healthy nodes, ensuring uninterrupted service for users.

\subsection{Llama-driven Checker}
SQL statements can bring many risks, including execution failure or irreversible impacts on the system. To address this problem, we have designed a new Llama driven syntax and security checker for StreamLink. These tools represent a significant advancement in enhancing the accuracy and security of SQL commands within our data engineering system.

The SQL syntax checker analyzes the structure and syntax of SQL commands generated by our system, ensuring that they adhere to the correct grammar and formatting rules. By validating the syntax of SQL commands, this tool significantly reduces the likelihood of errors that could arise from incorrect or malformed commands.Then the security checker plays a crucial role in mitigating potential risks associated with SQL injection attacks. By scrutinizing SQL commands for suspicious patterns or constructs that may indicate malicious intent, the security checker helps safeguard our system against unauthorized access, data breaches, and other security vulnerabilities.

Together, the SQL syntax checker and security checker strengthen the reliability and integrity of our data engineering system by minimizing the risk of errors and malicious activities. This proactive approach to SQL command validation not only enhances the overall quality of data processing but also instills confidence in the security posture of our system. It ensures the safe handling of sensitive information and protects against potential threats.


\section{Experiments}
In this section, we present the results of experiments conducted using StreamLink for data engineering, compared to traditional data systems. These experiments involve SQL generation reliability and malicious SQL interception evaluation.


\subsection{Generation Accuracy}

In our first experiment, we compared our proposed method to several existing approaches using the Spider\cite{yu2018spider} dataset, which consists of 10,181 questions and 5,693 unique complex SQL commands on 200 databases with multiple tables covering 138 different domains. Our goal was to evaluate the effectiveness of SQL generation, and we leveraged state-of-the-art LLMs and fine-tuning techniques to do so. The results showed that our method consistently outperformed the baseline methods in terms of SQL generation quality and accuracy.


We conduct experiments on Spider and compare our method with several baselines including:
\begin{itemize}
    \item Natural SQL\cite{gan2021natural}, a SQL intermediate representation (IR), enables existing models that do not support executable SQL generation to generate executable SQL queries.
    \item GRAPPA\cite{yu2020grappa}, a grammar-augmented pre-training framework for table semantic parsing.
    \item $S^2$SQL\cite{hui2022s}, injecting Syntax to question-Schema graph encoder for Text-to-SQL parsers, which effectively leverages the syntactic dependency information of questions in text-to-SQL to improve the performance.
    \item PICARD\cite{scholak2021picard}, a method for constraining auto-regressive decoders of language models through incremental parsing.
    \item RASAT\cite{qi2022rasat}, a Transformer-based seq2seq architecture augmented with relation-aware self-attention that could leverage a variety of relational structures.
    \item StruG\cite{deng2020structure}, structure-grounded pre-training framework (STRUG) for text-to-SQL that can effectively learn to capture text-table alignment based on a parallel text-table corpus.
    \item BERT\cite{devlin2018bert}, pre-training of deep bidirectional transformers for language understanding.
\end{itemize}

We demonstrate the exact match and execution accuracy between the baseline methods and our LLM-driven methods in Table \ref{tab:ex-em}.

\begin{table}[!ht]
    \centering
    \begin{tabular}{llll}
        \hline
        Approach & Exact Match & Accuracy \\
        \hline
        GRAPPA + RAT-SQL & 73.4 & - \\ 
        StruG + RAT-SQL & 72.6 & 74.9 \\ 
        BERT\_LARGE + RAT-SQL & 69.8 & 72.3 \\ 
        S2SQL + ELECTRA & 76.4 & - \\ 
        PICARD & 75.5 & 79.3 \\ 
        PICARD + RASAT & 75.3 & 80.5 \\ 
        RASAT & 72.6 & 76.6 \\
        T5-3B & 71.5 & 74.4 \\ 
        T5-3B + PICARD & 75.5 & 79.3 \\ 
        \hline
        SSQLG2-7B & \textbf{80.1} & \textbf{81.5} \\ 
        SSQLG2-13B & \textbf{81.9} & \textbf{82.7} \\ 
        SSQLG3-8B & \textbf{86.7} & \textbf{88.5} \\ 
        SSQLG3.1-8B & \textbf{86.9} & \textbf{89.7} \\ 
        \hline
    \end{tabular}
    \caption{Comparison of various models performance on spider dev-set for text-to-SQL, including Exact Match (EM) and Execution Accuracy (EA), the performance of StreamLink-SQL-Generator(SSQLG) is higher than that of Baseline. The best one is SSQLG3.1-8B which we fine-tuned on Llama-3.1-8B.}
    \label{tab:ex-em}
\end{table}

Instead of directly deploying off-the-shelf commercial or open-source LLMs, we hope to use domain knowledge to gain a StreamLink-dedicated model. The data in the table shows that our fine-tuned model has exceeded the baseline by over 10\% in both execution accuracy and exact match, achieving the effect of transferring a general language model to a specialized task. This provides the opportunity of using natural language interaction for StreamLink's users with different backgrounds. For instance, we can enable our lawyer collaborators to use natural language to perform specific patent analysis by saying ``tell me the top 10 most frequently appeared CPC by the assignee of Intel after 2009'' instead of manually writing a complex SQL command mentioned before.


These results highlight the effectiveness of our approach in addressing the challenges of SQL generation tasks, especially in complex and specialized domains with varying database schema. By outperforming existing methods on the Spider dataset, our method showcases its potential to significantly improve the efficiency and accuracy of SQL generation processes. This, in turn, can facilitate more effective data engineering and analysis workflows.

\subsection{Malicious SQL Interception}
In this experiment, we focused on evaluating the effectiveness of our SQL syntax checker and security checker based on Llama2. We used the SQL injection dataset \footnote{https://www.kaggle.com/datasets/syedsaqlainhussain/sql-injection-dataset} from Kaggle, which includes 30,595 SQL statements. Within this dataset, 19,258 were normal SQL, while 11,337 were malicious statements. We evaluated our LLM-based syntax and security checkers across different model sizes and model types. This dataset is representative of different SQL injections that occur in real-world scenarios, making it a solid testing ground for our tools.

Our evaluation focused on zero-shot conditions to simulate the checker's performance in situations where the specific dataset might not be feasible to train. This is common in organizations that need to adapt quickly to emerging threats without retraining models. We use recall and precision as metrics.

\begin{align}
    Recall = \frac{TP}{TP + FN} \\
    Precision = \frac{TP}{TP + FP} \\
    Escape = \frac{FN}{TP + FN} \\
    Misintercept = \frac{FP}{TN + FP}
\end{align}

Where
\begin{itemize}
    \item TP (True Positive) – Positive in the label, and predicted positive.
    \item FP (False Positive) – Negative in the label, but predicted positive.
    \item FN (False Negative) – Positive in the label, but predicted negative.
    \item FN (False Negative) – Negative in the label, and predicted negative.
\end{itemize}

After conducting multiple groups of random tests, we evaluate the effect of the model in the following table:

\begin{table}[!ht]
    \centering
    \begin{tabular}{lllll}
    \hline
        Approach & Precision & Recall & Escape & Misintercept \\ \hline
        SSQLC2: 7B & 76.54\% & 89.39\% & 10.61\% & 16.26\% \\ 
        SSQLC2: 70B & 74.2\% & 97.05\% & 2.95\% & 19.87\% \\ \hline
        SSQLC3: 8B & 79.31\% & 98.09\% & 1.91\% & 15.07\% \\ 
        SSQLC3: 70B & 80.01\% & 98.42\% & 1.58\% & 14.47\% \\ \hline
        SSQLC3.1: 8B & 71.7\% & 91.72\% & 8.28\% & 21.23\% \\ 
        SSQLC3.1: 70B & 90.52\% & 94.38\% & 5.62\% & 5.82\% \\ \hline
    \end{tabular}
    \caption{Test results of LLM of different sizes on malicious SQL data sets, we implement four tpyes of SQL checker based on Llama-2, Llama-3 and Llama-3.1, and show the test result of StreamLink-SQL-Checker (SSQLC in the table.)}
    \label{tab:recall-precision}
\end{table}

The data in Table \ref{tab:recall-precision} reflects the challenges posed by the Llama2 architecture, which, despite being effective, shows limitations in handling SQL interception compared to the more advanced Llama3 and Llama3.1 models. Specifically, the SSQLC2 series, based on Llama2, exhibits lower performance across most metrics. For instance, SSQLC2-70B achieves a recall of 97.05\%, which is impressive but still falls short of the results obtained with Llama3 and Llama3.1-based models. The precision of the SSQLC2 series also lags behind, highlighting that the older architecture and potentially outdated knowledge embedded in Llama2 lead to a higher rate of false positives, indicating a less reliable performance in real-world SQL injection detection.


The results for the Llama3.1-based models suggest that the training data and knowledge incorporated into this version may not have been as well-optimized for SQL interception as those in Llama3. The SSQLC3.1-8B model, for example, shows a noticeable drop in precision (71.7\%) compared to SSQLC3-8B (79.31\%), alongside a higher misintercept rate (21.23\% vs. 15.07\%). Although the SSQLC3.1-70B model does recover some ground, achieving a precision of 90.52\%, its performance inconsistencies relative to Llama3 indicate that Llama3.1 may not yet offer the same level of robustness for SQL attack detection.

Considering the balance between speed, accuracy, escape rate, and misintercept rate, the SSQLC3-8B model emerges as the most suitable choice for the StreamLink SQL Checker. It offers a strong recall rate of 98.09\% with a manageable precision of 79.31\%, all while maintaining a reasonable processing speed of 4 SQL statements per second. This model provides a well-rounded performance that meets the demands of real-time SQL injection detection while avoiding the significant speed drawbacks of the larger 70B models. The SSQLC-3-8B's combination of efficiency and effectiveness makes it the optimal solution for deployment in environments where both accuracy and speed are crucial.

\begin{figure}[htbp]
  \centering
  \includegraphics[width=\linewidth]{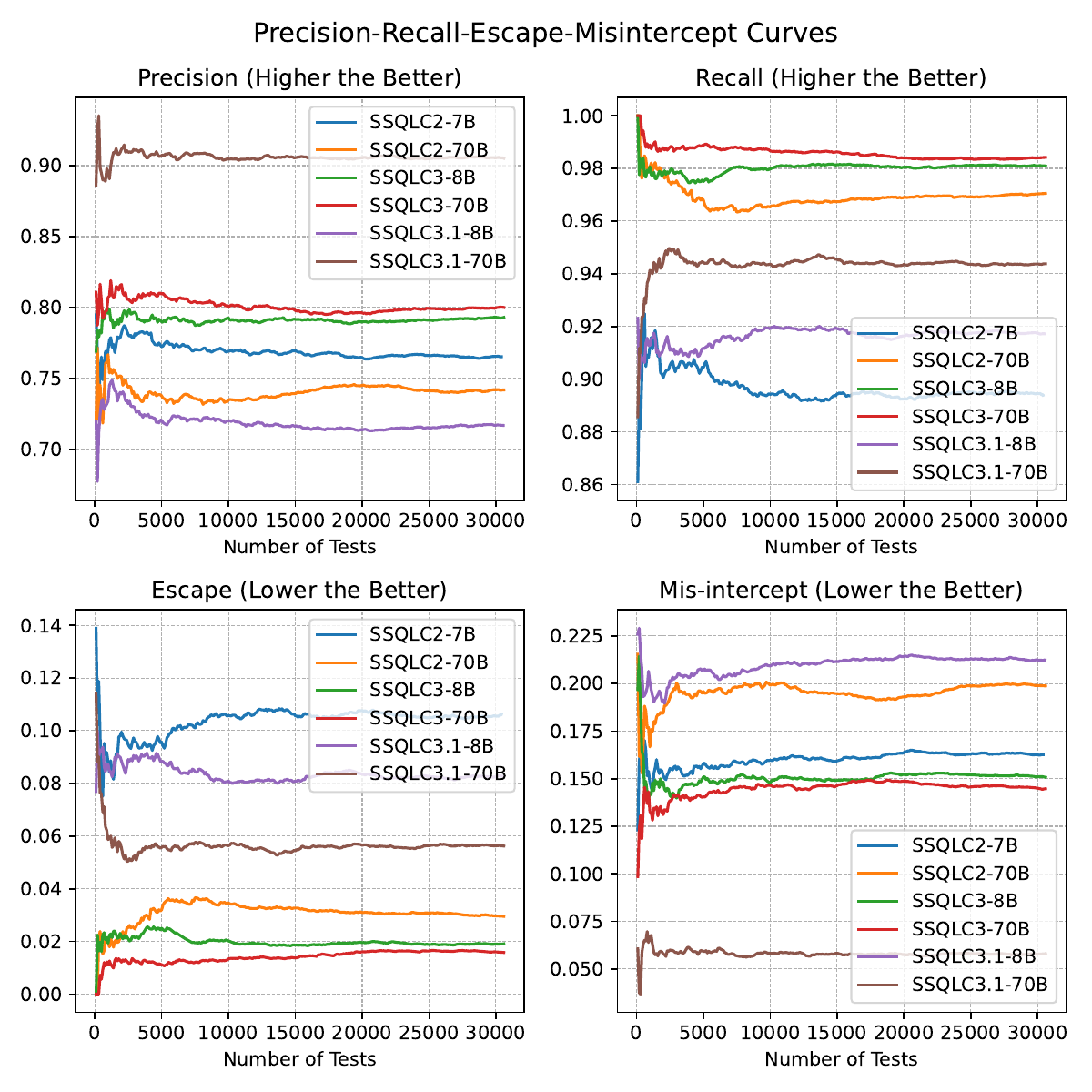}
  \caption{Malicious SQL interception analyzing on our LLM-based method}
  \label{fig:precision-recall}
\end{figure}

\begin{figure}[htbp]
  \centering
  \includegraphics[width=\linewidth]{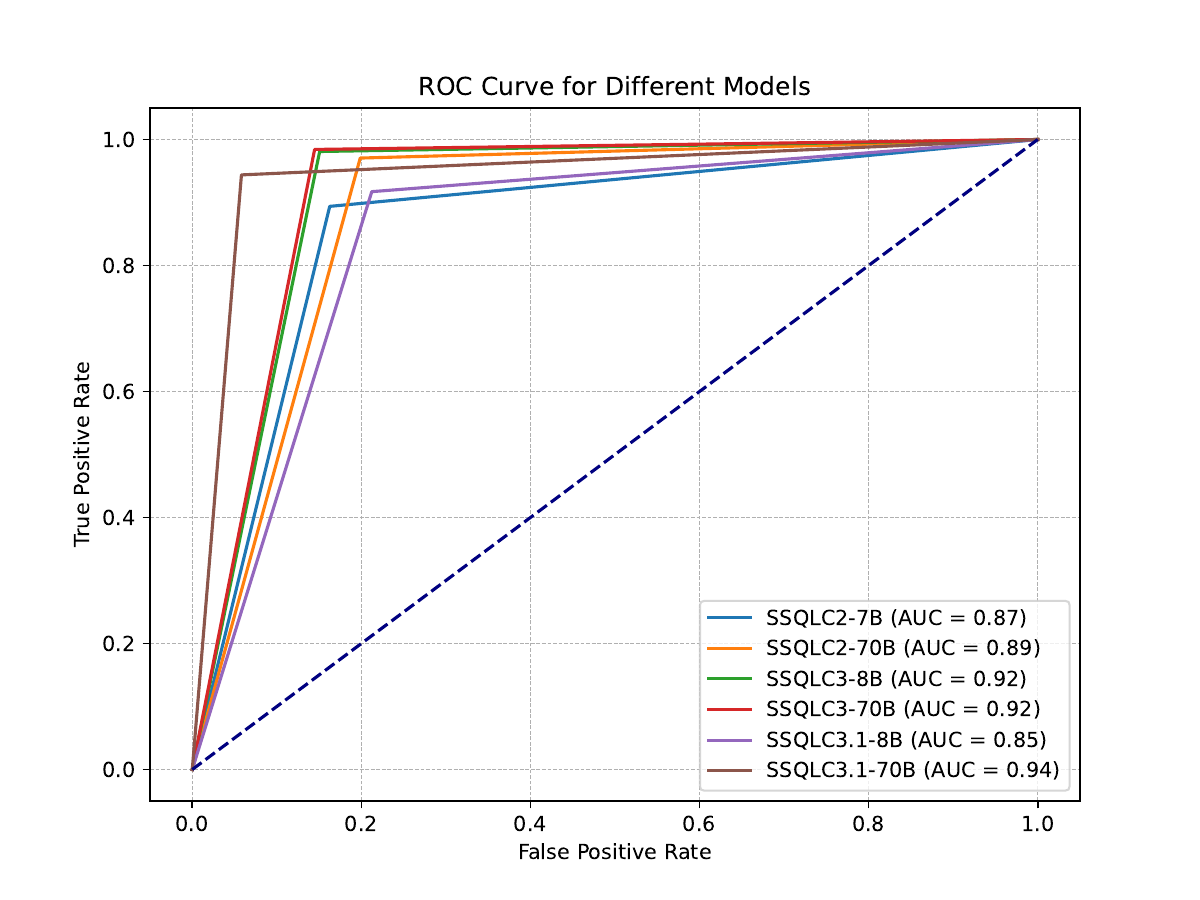}
  \caption{ROC Curve of our SSQLC methods, and the AUC of each method}
  \label{fig:roc}
\end{figure}
    
Figure \ref{fig:precision-recall} shows the test results obtained on sample sets of different sizes. When the sample size is less than 5000, the model's performance exhibits some fluctuations, which may be due to the uneven distribution of positive and negative samples in small samples. However, as the sample size increases from 5000 to 30000, the distribution of positive and negative labels gradually approaches normal distribution, and the model demonstrates excellent stability. 

The results of the experiment were highly encouraging, Figure \ref{fig:roc} indicates that our interceptors provide robust protection against malicious SQL commands. By effectively identifying and blocking malicious actions, our system ensures the stable operation of the server, safeguarding against potential disruptions and data breaches. This demonstrates the critical role of our SQL syntax checker and security checker in fortifying the system's defenses against malicious attacks and ensuring the reliability and security of data processing operations.

\section{Conclusion}

In conclusion, we present a comprehensive exploration of StreamLink, an innovative data engineering system empowered by cutting-edge technologies such as LLMs, distributed computing frameworks, and advanced security mechanisms. Through a series of experiments and evaluations, we have demonstrated the effectiveness, efficiency, and security of StreamLink in various data engineering tasks.

StreamLink's scalable architecture, where all services interact via an Application Programming Interface (API) and Remote Procedure Call (RPC) structure, provides rich extensibility for the system. Our evaluations have shown that StreamLink significantly outperforms existing solutions in specific domains, making it a robust and transformative tool for data engineering. The integration of advanced algorithms and a focus on scalability and security positions StreamLink as a solution for organizations looking to enhance their data workflows while maintaining high standards of privacy and security.

Overall, we think StreamLink represents a significant advancement in data engineering technology, offering unparalleled capabilities in terms of efficiency, scalability, and security. Its innovative use of LLMs, combined with distributed computing frameworks and advanced security mechanisms, positions StreamLink as a transformative solution for organizations seeking to enhance their data engineering workflows.

\clearpage

\bibliographystyle{ACM-Reference-Format}

\end{document}